\documentclass[fleqn,10pt]{wlscirep}
\usepackage[utf8]{inputenc}
\usepackage[T1]{fontenc}
\usepackage{siunitx}

\usepackage{xcolor}
\usepackage[normalem]{ulem} 

\definecolor{revadd}{RGB}{0,140,90}     
\definecolor{revdel}{RGB}{200,40,40}    
\definecolor{revcom}{RGB}{140,0,180}    

\newif\ifrev
\revtrue          

\ifrev
  
  \newcommand{\del}[1]{\textcolor{revdel}{\sout{#1}}}
  
  \newcommand{\com}[1]{\textcolor{revcom}{\textbf{[}#1\textbf{]}}}
\else
  
  \newcommand{\del}[1]{}
  
  \newcommand{\com}[1]{}
\fi

\title{Sizing the Universe with DESI Galaxy Sizes: Plain Fundamentals of Fundamental-Plane Lensing}

\author[1,*]{Kun Xu}
\author[2,3]{Jiaqi Wang}
\author[1]{Ravi K. Sheth}

\author[4]{J.~Aguilar}
\author[5]{S.~Ahlen}
\author[6]{F.~Beutler}
\author[7,8]{D.~Bianchi}
\author[9]{D.~Brooks}
\author[10,11]{A.~Carnero Rosell}
\author[4]{T.~Claybaugh}
\author[12]{A.~de la Macorra}
\author[9]{P.~Doel}
\author[13,14]{J.~E.~Forero-Romero}
\author[15,16,17]{E.~Gaztañaga}
\author[18]{G.~Gutierrez}
\author[19,20,21]{K.~Honscheid}
\author[22]{C.~Howlett}
\author[23]{M.~Ishak}
\author[24]{J.~Jimenez}
\author[25]{S.~Juneau}
\author[26]{R.~Kehoe}
\author[27]{D.~Kirkby}
\author[9]{O.~Lahav}
\author[4]{M.~Landriau}
\author[28]{L.~Le~Guillou}
\author[29,24]{M.~Manera}
\author[25]{A.~Meisner}
\author[30,24]{R.~Miquel}
\author[31]{J.~Moustakas}
\author[16]{S.~Nadathur}
\author[32]{J.~A.~Newman}
\author[33,34,35]{W.~J.~Percival}
\author[36]{F.~Prada}
\author[37]{I.~P\'erez-R\`afols}
\author[38]{C.~Ravoux}
\author[39]{G.~Rossi}
\author[40]{R.~Ruggeri}
\author[41]{E.~Sanchez}
\author[42]{C.~Saulder}
\author[4]{D.~Schlegel}
\author[43,44]{M.~Schubnell}
\author[45]{H.~Seo}
\author[4]{J.~Silber}
\author[17,11]{M.~Siudek}
\author[44]{G.~Tarl\'{e}}
\author[25]{B.~A.~Weaver}
\author[46]{H.~Zou}

\affil[1]{Center for Particle Cosmology, Department of Physics and Astronomy, University of Pennsylvania, Philadelphia, PA 19104, USA}
\affil[2]{State Key Laboratory of Dark Matter Physics, Tsung-Dao Lee Institute \& School of Physics and Astronomy, Shanghai Jiao Tong University, Shanghai 201210, China}
\affil[3]{Institute for Computational Cosmology, Department of Physics, Durham University, South Road, Durham, DH1 3LE, UK}

\affil[4]{Lawrence Berkeley National Laboratory, 1 Cyclotron Road, Berkeley, CA 94720, USA}
\affil[5]{Department of Physics, Boston University, 590 Commonwealth Avenue, Boston, MA 02215 USA}
\affil[6]{Institute for Astronomy, University of Edinburgh, Royal Observatory, Blackford Hill, Edinburgh EH9 3HJ, UK}
\affil[7]{Dipartimento di Fisica ``Aldo Pontremoli'', Universit\`a degli Studi di Milano, Via Celoria 16, I-20133 Milano, Italy}
\affil[8]{INAF-Osservatorio Astronomico di Brera, Via Brera 28, 20122 Milano, Italy}
\affil[9]{Department of Physics \& Astronomy, University College London, Gower Street, London, WC1E 6BT, UK}
\affil[10]{Departamento de Astrof\'{\i}sica, Universidad de La Laguna (ULL), E-38206, La Laguna, Tenerife, Spain}
\affil[11]{Instituto de Astrof\'{\i}sica de Canarias, C/ V\'{\i}a L\'{a}ctea, s/n, E-38205 La Laguna, Tenerife, Spain}
\affil[12]{Instituto de F\'{\i}sica, Universidad Nacional Aut\'{o}noma de M\'{e}xico, Circuito de la Investigaci\'{o}n Cient\'{\i}fica, Ciudad Universitaria, Cd. de M\'{e}xico C.~P.~04510, M\'{e}xico}
\affil[13]{Departamento de F\'isica, Universidad de los Andes, Cra. 1 No. 18A-10, Edificio Ip, CP 111711, Bogot\'a, Colombia}
\affil[14]{Observatorio Astron\'omico, Universidad de los Andes, Cra. 1 No. 18A-10, Edificio H, CP 111711 Bogot\'a, Colombia}
\affil[15]{Institut d'Estudis Espacials de Catalunya (IEEC), c/ Esteve Terradas 1, Edifici RDIT, Campus PMT-UPC, 08860 Castelldefels, Spain}
\affil[16]{Institute of Cosmology and Gravitation, University of Portsmouth, Dennis Sciama Building, Portsmouth, PO1 3FX, UK}
\affil[17]{Institute of Space Sciences, ICE-CSIC, Campus UAB, Carrer de Can Magrans s/n, 08913 Bellaterra, Barcelona, Spain}
\affil[18]{Fermi National Accelerator Laboratory, PO Box 500, Batavia, IL 60510, USA}
\affil[19]{Center for Cosmology and AstroParticle Physics, The Ohio State University, 191 West Woodruff Avenue, Columbus, OH 43210, USA}
\affil[20]{Department of Physics, The Ohio State University, 191 West Woodruff Avenue, Columbus, OH 43210, USA}
\affil[21]{The Ohio State University, Columbus, 43210 OH, USA}
\affil[22]{School of Mathematics and Physics, University of Queensland, Brisbane, QLD 4072, Australia}
\affil[23]{Department of Physics, The University of Texas at Dallas, 800 W. Campbell Rd., Richardson, TX 75080, USA}
\affil[24]{Institut de F\'{i}sica d’Altes Energies (IFAE), The Barcelona Institute of Science and Technology, Edifici Cn, Campus UAB, 08193, Bellaterra (Barcelona), Spain}
\affil[25]{NSF NOIRLab, 950 N. Cherry Ave., Tucson, AZ 85719, USA}
\affil[26]{Department of Physics, Southern Methodist University, 3215 Daniel Avenue, Dallas, TX 75275, USA}
\affil[27]{Department of Physics and Astronomy, University of California, Irvine, 92697, USA}
\affil[28]{Sorbonne Universit\'{e}, CNRS/IN2P3, Laboratoire de Physique Nucl\'{e}aire et de Hautes Energies (LPNHE), FR-75005 Paris, France}
\affil[29]{Departament de F\'{i}sica, Serra H\'{u}nter, Universitat Aut\`{o}noma de Barcelona, 08193 Bellaterra (Barcelona), Spain}
\affil[30]{Instituci\'{o} Catalana de Recerca i Estudis Avan\c{c}ats, Passeig de Llu\'{\i}s Companys, 23, 08010 Barcelona, Spain}
\affil[31]{Department of Physics and Astronomy, Siena University, 515 Loudon Road, Loudonville, NY 12211, USA}
\affil[32]{Department of Physics \& Astronomy and Pittsburgh Particle Physics, Astrophysics, and Cosmology Center (PITT PACC), University of Pittsburgh, 3941 O'Hara Street, Pittsburgh, PA 15260, USA}
\affil[33]{Department of Physics and Astronomy, University of Waterloo, 200 University Ave W, Waterloo, ON N2L 3G1, Canada}
\affil[34]{Perimeter Institute for Theoretical Physics, 31 Caroline St. North, Waterloo, ON N2L 2Y5, Canada}
\affil[35]{Waterloo Centre for Astrophysics, University of Waterloo, 200 University Ave W, Waterloo, ON N2L 3G1, Canada}
\affil[36]{Instituto de Astrof\'{i}sica de Andaluc\'{i}a (CSIC), Glorieta de la Astronom\'{i}a, s/n, E-18008 Granada, Spain}
\affil[37]{Departament de F\'isica, EEBE, Universitat Polit\`ecnica de Catalunya, c/Eduard Maristany 10, 08930 Barcelona, Spain}
\affil[38]{Universit\'{e} Clermont-Auvergne, CNRS, LPCA, 63000 Clermont-Ferrand, France}
\affil[39]{Department of Physics and Astronomy, Sejong University, 209 Neungdong-ro, Gwangjin-gu, Seoul 05006, Republic of Korea}
\affil[40]{Queensland University of Technology, School of Chemistry \& Physics, George St, Brisbane 4001, Australia}
\affil[41]{CIEMAT, Avenida Complutense 40, E-28040 Madrid, Spain}
\affil[42]{Max Planck Institute for Extraterrestrial Physics, Gie\ss enbachstra\ss e 1, 85748 Garching, Germany}
\affil[43]{Department of Physics, University of Michigan, 450 Church Street, Ann Arbor, MI 48109, USA}
\affil[44]{University of Michigan, 500 S. State Street, Ann Arbor, MI 48109, USA}
\affil[45]{Department of Physics \& Astronomy, Ohio University, 139 University Terrace, Athens, OH 45701, USA}
\affil[46]{National Astronomical Observatories, Chinese Academy of Sciences, A20 Datun Road, Chaoyang District, Beijing, 100101, P.~R.~China}

\affil[*]{kunxu@sas.upenn.edu}

\begin{abstract}
Weak gravitational lensing provides a powerful way to map cosmic structure, but most current measurements rely on galaxy shape distortions from deep imaging surveys and are affected by systematics such as intrinsic alignments, photometric-redshift uncertainties and shape-measurement biases. Here we present a spectroscopic galaxy--galaxy lensing magnification measurement using Fundamental-Plane (FP) size residuals, $\delta_r\equiv\Delta\log_{10}R_\mathrm{e}$, from 3.26 million DESI luminous red galaxy (LRG) sources behind DESI Bright Galaxy Survey lenses. The FP-like relation predicts the intrinsic sizes of LRGs from lensing-invariant quantities, including velocity dispersion $\sigma_0$ and surface brightness $I_\mathrm{e}$, with a scatter of about 0.06-0.07 dex. Lensing magnifies LRG sizes, giving the direct convergence response $\delta_r(\kappa)=\kappa/\ln10$, but we show that the full lensing response is modified by magnification bias, because fitting $R_\mathrm{e}$ with $I_\mathrm{e}$ inevitably induces a magnitude dependence in $\bar{\delta}_r(m)$. After calibrating this response, we recover surface-density profiles with uncertainties comparable to those from individual Stage-III shear surveys using 5--20 million higher-redshift sources. The corresponding excess surface-density profiles agree with shear-based measurements. We further show that the estimator is robust to size-measurement uncertainties, with a convergence multiplicative bias only $\simeq -0.2$ times the size bias. FP lensing therefore provides a clean, spectroscopic and complementary probe of cosmic structure. 
\end{abstract}
\begin{document}

\flushbottom
\maketitle

\section*{Main}
As light from distant sources propagates through the inhomogeneous Universe, it is deflected by the gravitational potential of foreground large-scale structure, producing gravitational lensing \cite{2001PhR...340..291B}. Gravitational lensing therefore provides a direct probe of the matter distribution. It changes the observed images of background sources in two main ways: shear anisotropically distorts their apparent shapes, whereas magnification changes their apparent angular sizes while conserving surface brightness. Current measurements of cosmic structure with gravitational lensing rely primarily on the weak distortion of galaxy shapes by shear, commonly referred to as weak lensing (WL) \cite{1993ApJ...404..441K}.

Although theoretically elegant, shape-based WL measurements suffer from a range of observational and modelling systematics, including intrinsic alignments (IA), photometric-redshift uncertainties and shape-measurement biases \cite{2018ARA&A..56..393M}. Relying on shear measurements from galaxy shapes, current Stage-III surveys, including the Dark Energy Survey (DES), the Kilo-Degree Survey (KiDS) and the Hyper Suprime-Cam Subaru Strategic Program (HSC), have measured the late-time matter clustering amplitude, commonly parametrized by $S_8\equiv\sigma_8(\Omega_{\rm m}/0.3)^{0.5}$, with percent-level precision \cite{2026arXiv260114559D,2025A&A...703A.158W,2023PhRvD.108l3521S}. Among these measurements, some find an approximately $2\sigma$ tension with the prediction of the $\Lambda$ cold dark matter ($\Lambda$CDM) model inferred from Planck\cite{2020A&A...641A...6P} cosmic microwave background (CMB) measurements, while others find results consistent with Planck \cite{2026PDU....5202286P}. 

It is therefore crucial to develop weak-lensing measurements that are independent of galaxy-shape shear, both to provide complementary information for tighter constraints and to test consistency under different systematic uncertainties. Recent efforts include kinematic lensing as an alternative route to measuring shear\cite{2013arXiv1311.1489H,2025MNRAS.540.2877P}, as well as magnification measurements based on galaxy number counts\cite{2005ApJ...633..589S,2010MNRAS.405.1025M,2025PhRvD.111j3540S}, fluxes\cite{2010MNRAS.405.1025M,2024ApJ...973..102X} and sizes\cite{2006ApJ...648L..17B,2014ApJ...780L..16H,2015MNRAS.454..478J}, although their accuracy is not yet comparable to that of galaxy-shear measurements. 

Among these approaches, Fundamental-Plane (FP) lensing with galaxy sizes is particularly promising \cite{2006ApJ...648L..17B,2014ApJ...780L..16H,2015MNRAS.454..478J}. The key idea is to predict the intrinsic galaxy size from lensing-invariant quantities. For early-type galaxies, the FP provides a tight empirical relation \cite{1987ApJ...313...42D,1987ApJ...313...59D},
\begin{equation}
\log_{10} R_\mathrm{e}
= a \log_{10} \sigma_0 + b \log_{10} I_\mathrm{e} + c\, ,
\end{equation}
where the effective radius $R_\mathrm{e}$ is inferred from the central velocity dispersion $\sigma_0$ and the mean surface brightness $I_\mathrm{e}$ within $R_\mathrm{e}$, both of which are conserved under lensing. The intrinsic scatter of this relation is usually smaller than 0.1 dex\cite{2003AJ....125.1866B,2025arXiv251203226R}, substantially smaller than the shape noise in shear measurements. Magnification $\mu$ can therefore be measured from the size offset $\delta_r\equiv\Delta\log_{10}R_\mathrm{e}$ between the observed size and the FP-predicted intrinsic size. Since lensing changes angular size as $R_\mathrm{e}\rightarrow \mu^{1/2}R_\mathrm{e}$, the expected response is $\delta_r=\log_{10}\mu/2$. In the weak-lensing limit, this gives an estimate of the convergence, $\delta_r\simeq\kappa/\ln 10$.

In this paper, we present a galaxy-galaxy lensing (GGL) measurement based on the sizes of luminous red galaxies (LRGs) from the Dark Energy Spectroscopic Instrument (DESI) Data Release~2 (DR2), using three Bright Galaxy Survey (BGS) Bright lens samples from DESI Data Release~1 (DR1). We do not use the DR2 BGS lens sample because it is protected for the DESI lensing key paper. We show that previous FP-lensing studies \cite{2014ApJ...780L..16H,2015MNRAS.454..478J} did not fully account for the correct lensing response of the FP residual. With the correct response included, our FP-based GGL measurements achieve a precision comparable to shear-based GGL measurements using source galaxies from a single Stage-III imaging survey\cite{2025arXiv250621677H}. The FP- and shear-based measurements are consistent within the uncertainties. We emphasize that both the lens and source samples used here are spectroscopic, so the measurement is not affected by photometric-redshift uncertainties. Moreover, owing to the accurate spectroscopic redshifts, lenses and sources can be cleanly separated in redshift, making intrinsic correlations of FP residuals, analogous to IA in shear measurements, negligible for our GGL analysis. We also note that the FP has already been used as a cosmological probe with DESI, for example through peculiar-velocity measurements~\cite{2025MNRAS.539.3627S,2025arXiv251203226R}.

In Figure~\ref{fig:dis}(a,b), we show the angular distributions of the LRG source sample and the BGS lens samples. The overlapping region, covering approximately $7264~\mathrm{deg}^2$, is colour-coded by angular number density. The BGS lens samples are constructed in three redshift bins, $[0.1,0.2]$, $[0.2,0.3]$ and $[0.3,0.4]$, with $r$-band absolute-magnitude cuts of $M_r<-19.5$, $M_r<-20.5$ and $M_r<-21.0$, respectively \cite{2024MNRAS.533..589Y,2025arXiv250621677H}. The LRG source sample spans $0.4<z<1.1$. The redshift distributions are shown in Figure~\ref{fig:dis}(c), where the three lens samples are denoted BGS1, BGS2 and BGS3. After restricting to the overlap region, the three lens samples contain approximately 739k, 662k and 424k galaxies, respectively, and the source sample contains $3.26$ million LRGs. For BGS3 lenses, we further require LRG sources to satisfy $z>0.5$.

To measure $\delta_r$ for LRGs, we fit an extended FP model, denoted the ``quadIe+ML'' model, which includes a quadratic dependence on $I_\mathrm{e}$ and a stellar mass-to-light ratio term:
\begin{equation}
\log_{10} R_\mathrm{e}
= a \log_{10} \sigma_0
+ b_1 \log_{10} I_\mathrm{e}
+ b_2 \left(\log_{10} I_\mathrm{e}\right)^2
+ d\log_{10}(M_*/L)
+ c\, .
\end{equation}
All quantities on the right-hand side are conserved under lensing. We fit this relation in 14 narrow redshift bins and linearly interpolate the best-fitting parameters across the full redshift range. In Figure~\ref{fig:FP_fitting}, we compare the observed $R_\mathrm{e}$ with the model predictions and show the mean and scatter of the residuals for both the standard FP and the ``quadIe+ML'' model. The ``quadIe+ML'' model provides a better description of galaxies with the largest and smallest $R_\mathrm{e}$, and gives a slightly smaller residual scatter of $\sigma_\mathrm{quad}=0.06\,\mathrm{dex}$. The $\left(\log_{10} I_\mathrm{e}\right)^2$ term is introduced purely empirically, as a quadratic term is the most natural extension of the standard FP. The origin of this term, and whether it reflects a physical effect or arises from selection effects, is not investigated in this work, as it does not affect the lensing measurements. We then define the FP residual $\delta_r$ as the difference between the observed $\log_{10}R_\mathrm{e}$ and the ``quadIe+ML'' prediction. 

Moreover, only 1.37 million LRGs have measured $\sigma_0$, while 1.89 million do not. We fit the FP-like relations using only galaxies with measured $\sigma_0$. To include the remaining galaxies, we assume that they follow the same ``quadIe+ML'' relation in each redshift bin, replacing their individual velocity dispersions with a fitted mean $\sigma_0$ for the no-$\sigma_0$ subsample. The scatter in $\log_{10}R_\mathrm{e}$ around this relation then represents the size-prediction scatter using only $I_\mathrm{e}$ and $M_*/L$. This scatter increases only mildly to $\sigma_\mathrm{quad}=0.07\,\mathrm{dex}$, reflecting the relatively small intrinsic variation in $\sigma_0$ among LRGs.

To infer the convergence field from $\delta_r$, the lensing response of the FP residual must be specified. If lensing only changed the angular sizes of a fixed set of LRGs, the response would be the ideal size response, $\delta_r\simeq\kappa/\ln 10$, as assumed above and in previous FP-lensing studies \cite{2006ApJ...648L..17B,2014ApJ...780L..16H,2015MNRAS.454..478J}. In a real selected sample, however, magnification bias is unavoidable: lensing changes the apparent magnitudes and sizes of galaxies, and therefore changes which objects enter the sample near the selection boundaries.

\begin{figure}[ht]
\centering
\includegraphics[width=0.8\linewidth]{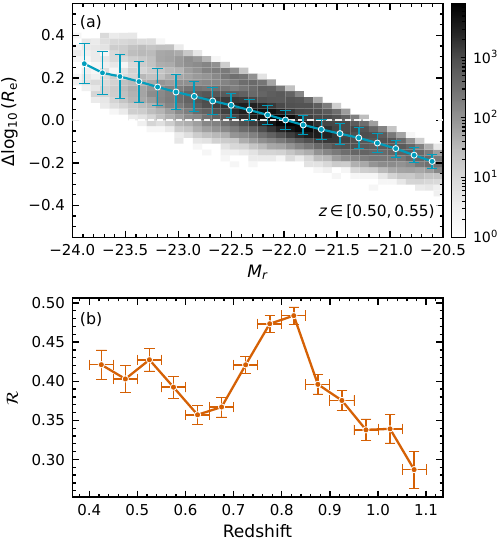}
\caption{
Response of the FP-like size residual to magnification.
(a) Dependence of the size residual $\Delta\log_{10}(R_\mathrm{e})$ on $r$-band absolute magnitude for one representative redshift bin, $0.50<z<0.55$. The grey-scale histogram shows the two-dimensional galaxy distribution, and the cyan points show the binned mean residual with scatter. The residuals show a clear magnitude dependence, indicating that changes in the source selection under magnification can modify the effective response of the FP-size estimator.  
(b) Effective lensing response $\mathcal{R}$ as a function of source redshift. The response is measured by applying a small test convergence, $\kappa=-0.003$, to the source sample and re-evaluating the mean FP residual in each redshift bin. The error bars are estimated from jackknife resampling.
}
\label{fig:response}
\end{figure}

This affects the FP residual because the surface-brightness variable itself couples magnitude and size, and therefore inevitably generates a magnitude dependence in the resulting $\delta_r$. By definition, at fixed surface brightness $I_\mathrm{e}$, $\partial \log_{10}R_\mathrm{e}/\partial M=-0.2$, where $M$ is the absolute magnitude. If the FP observables $\mathbf{x}=(\sigma_0,I_\mathrm{e},M_*/L,\ldots)$ are held fixed, the FP-predicted size is also fixed, so the residual has the same local dependence, $\partial\delta_r/\partial M=-0.2$. After marginalizing over the distribution $p(\mathbf{x}|M)$, this becomes a mean trend $d\bar{\delta}_r/dM$, which is generally smaller that 0.2 in amplitude but remains non-zero.  Figure~\ref{fig:response}(a) shows the dependence of $\bar{\delta}_r$ on $M_r$ of LRGs for one representative redshift bin. The measured slope is approximately $-0.095$, consistent with the argument above. 

As a result, galaxies that enter or leave the sample under magnification have, in general, a mean residual different from that of the original sample, $\bar{\delta}_r(m_{\rm lim})-\langle\delta_r\rangle\neq0$. This selection-induced change in the mean intrinsic residual modifies the response of the stacked FP residual to $\kappa$. The conversion from $\delta_r$ to convergence is therefore not simply the ideal size-only response, but instead takes the form $\langle\delta_r\rangle(\kappa)=\mathcal{R}\kappa/\ln 10$. For example, for a magnitude-limited selection,
\begin{equation}
\mathcal{R}=1+5\ln 10\,s(m_\mathrm{lim})\left[\bar{\delta}_r(m_\mathrm{lim})-\langle\delta_r\rangle(0)\right],
\end{equation}
where $s(m_\mathrm{lim})$ is the logarithmic slope of the cumulative number counts at the limiting magnitude. For the DESI LRG sample, the response can be more complex because the selection involves multiple colour and fibre magnitude cuts rather than a single limiting magnitude. A more detailed explanation of the response is given in the \textbf{Methods}.

Fortunately, the magnification bias response can be determined directly from the sample itself.  We then apply a small demagnification, $d\kappa=-0.003$, to the catalogue and measure the corresponding response factor $\mathcal{R}$. Figure~\ref{fig:response}(b) shows $\mathcal{R}$ in different redshift bins, with uncertainties estimated by jackknife resampling. We find $\mathcal{R}\simeq0.3$--$0.5$ for DESI LRGs. 

\begin{figure}[ht]
\centering
\includegraphics[width=0.8\linewidth]{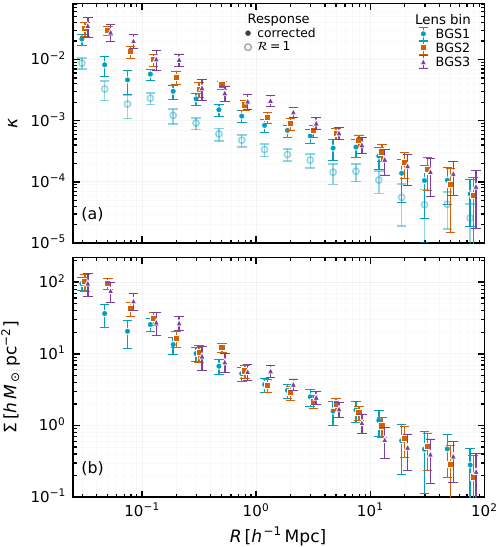}
\caption{
FP-based galaxy--galaxy lensing measurements for the three DESI BGS lens samples.
(a) Convergence profiles $\kappa(R)$ measured from the FP-size residuals of DESI DR2 LRG sources around the BGS1, BGS2 and BGS3 lens samples. For BGS1, filled symbols show the fiducial measurement with the selection-response correction applied, while open symbols show the result obtained when this correction is neglected, i.e. when $\mathcal{R}=1$ is assumed. The other lens samples show similar differences.
(b) Corresponding comoving surface-density profiles $\Sigma(R)$ for the fiducial response-corrected measurements, obtained using the pair-dependent comoving critical surface density.
All profiles are measured as a function of projected comoving separation $R$. Error bars show jackknife uncertainties estimated from 200 sky regions.
}
\label{fig:GGL}
\end{figure}

In Figure~\ref{fig:GGL}, we show the estimated convergence $\kappa(R)$ and surface-density $\Sigma(R)$ profiles around the BGS1, BGS2 and BGS3 lens samples. Details of the measurement procedure are given in the \textbf{Methods}. We detect a clear FP-lensing signal for all three lens samples over a wide radial range, $0.025\,h^{-1}\mathrm{Mpc}<R<100\,h^{-1}\mathrm{Mpc}$. As a consistency check, we also measure $\Sigma(R)$ using LRG sources split into redshift subsamples, as shown in Extended Data Figure~\ref{fig:GGL_Sigma_sourcebins}. The profiles measured from different source-redshift bins are consistent with each other for all lens samples, providing an internal validation of the measurement.

\begin{figure}[ht]
\centering
\includegraphics[width=0.6\linewidth]{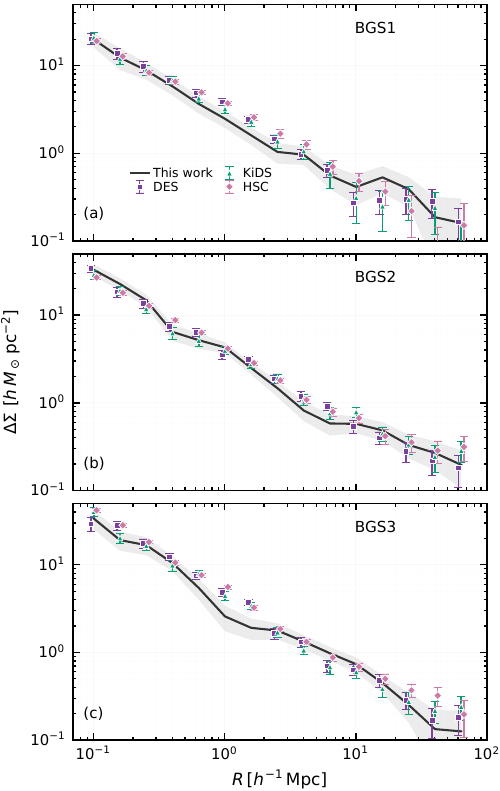}
\caption{
Comparison between FP-based and shear-based galaxy--galaxy lensing measurements.
Panels (a), (b) and (c) show the excess surface-density profiles $\Delta\Sigma(R)$ for the BGS1, BGS2 and BGS3 lens samples, respectively. The black curves show the FP lensing measurements from this work, obtained by converting the measured $\Sigma(R)$ profiles to $\Delta\Sigma(R)$. The grey bands show the corresponding jackknife uncertainties. The coloured points show shear-based measurements using the same DESI lens samples with DES Y3, KiDS-1000 and HSC Y3 shear catalogues.
}
\label{fig:compare_to_shear}
\end{figure}

We compare our measurements with existing shear-based GGL results using the same DESI lens samples by integrating $\Sigma(R)$ to obtain the excess surface density $\Delta\Sigma(R)$. In Figure~\ref{fig:compare_to_shear}, we compare the derived $\Delta\Sigma(R)$ with measurements from DES Y3 (overlap area $662\,\mathrm{deg}^2$, $n_{\rm eff}=5.59\,\mathrm{arcmin}^{-2}$), KiDS-1000 ($448\,\mathrm{deg}^2$, $n_{\rm eff}=6.17\,\mathrm{arcmin}^{-2}$), and HSC Y3 ($453\,\mathrm{deg}^2$, $n_{\rm eff}=19.9\,\mathrm{arcmin}^{-2}$) shear catalogues~\cite{2025arXiv250621677H}. Our results are consistent with these shear measurements, with uncertainties comparable to those from an individual shear survey, albeit slightly larger. This is notable because our measurement uses only 3.26 million LRG sources, compared with the 5--20 million overlapping sources used in the shear measurements, and has lower lensing efficiency owing to the lower source redshifts. These results demonstrate that FP lensing can provide a promising, independent probe of cosmic structure. We note that, with $\mathcal{R}\simeq0.4$, the effective convergence noise is
$0.06\times\ln 10/0.4\simeq0.35$, comparable to the shape noise in shear measurements. Nevertheless, this suppression does not remove the statistical advantage of FP lensing, since shear estimators also have a response factor of around $0.6$, although its value varies among surveys, and measurement uncertainties in both galaxy sizes and shapes further contribute to the effective noise. This helps explain why our FP measurements can still achieve comparable precision with fewer sources. We further expect that optimized source weighting could increase the effective $\mathcal{R}$.

Another advantage of FP lensing is its robustness to size-measurement bias. For the standard FP fit to LRGs, with $b_\mathrm{obs}\simeq-0.6$, a multiplicative bias in the size measurement, $\log_{10}R_\mathrm{e}'=(1+m_r)\log_{10}R_\mathrm{e}+c_r$, induces only a small multiplicative bias $\kappa'=(1+\mathcal{M})\kappa$ in the inferred convergence, with
\begin{equation}
    \mathcal{M}\simeq-0.2m_r\,.
    \label{eq:size2kappaBias}
\end{equation}
A detailed derivation is given in the \textbf{Methods}. As a direct test, we repeat the $\Sigma(R)$ measurement around BGS1 after applying a deliberately large size rescaling, $\log_{10}R_\mathrm{e}'=0.5\log_{10}R_\mathrm{e}+0.35$, and refitting the ``quadIe+ML'' model. As shown in Extended Data Figure~\ref{fig:Re_multibias}. the rescaled measurement remains consistent with the fiducial result within the uncertainties, with only a slightly higher amplitude, confirming the expected weak sensitivity to size-measurement bias.

In conclusion, we have presented an FP-based GGL measurement using galaxy sizes from only 3.26 million DESI DR2 LRG sources, achieving uncertainties comparable to those from an individual Stage-III shear survey with 5--20 million sources. After calibrating the correct response to selection-induced magnification bias, which has not been fully accounted for in previous studies\cite{2006ApJ...648L..17B,2014ApJ...780L..16H,2015MNRAS.454..478J}, we obtain a matter-density measurement consistent with shear-based measurements. We note that this selection effect may also be relevant for lensing measurements based on Tully--Fisher relation\cite{1977A&A....54..661T}, such as kinematic lensing\cite{2013arXiv1311.1489H,2025MNRAS.540.2877P}. Because our measurement uses only spectroscopic samples, it avoids biases from photometric-redshift uncertainties and reduces contamination from intrinsic size
and/or shape correlations associated with physically associated source--lens pairs that complicate shear-based lensing analyses. 
Indeed, FP lensing may one day provide a diagnostic of the impact of IA on shear-based analyses.
FP lensing is also relatively robust to size multiplicative bias (c.f. Eq.~\ref{eq:size2kappaBias} for DESI LRGs).  
Finally, we note that the LRG sample used here was not optimized for FP lensing. An optimized source selection or weighting scheme could further enhance the lensing response $\mathcal{R}$ and improve the signal-to-noise ratio, which will be important to investigate in future work. 

Although the DESI LRG spectroscopic sample provides a clean FP-lensing signal with precision comparable to a Stage-III shear survey, improving the accuracy by adding more high-redshift sources can be challenging. This is partly because measuring $\sigma_0$ is harder than measuring $z$. If $z$ is known but $\sigma_0$ is not, one can use the $R_\mathrm{e}$--$I_\mathrm{e}$ relation rather than the full FP; the resulting GGL estimate will be noisier, but not otherwise biased. This could be improved by replacing $\sigma_0$ with other lensing-invariant observables that correlate with $\sigma_0$, such as rest-frame colours \cite{2005AJ....129...61B} or the concentration of the surface-brightness profile \cite{2014ApJ...780L..16H}.

Even so, achieving the statistical power of Stage-IV shear measurements, with billions of sources, is more challenging, simply because of the limited number of spectroscopic galaxies expected in the foreseeable future. However, FP lensing can also be measured in purely photometric surveys, although one must first account for the correlated effects of photometric-redshift errors on the estimated $R_\mathrm{e}$, $I_\mathrm{e}$ and other galaxy properties. Previous work suggests that this is not an insurmountable challenge \cite{2010MNRAS.401..666R,2010MNRAS.403.2137S}. In addition, photometric-redshift errors will affect the modelling of the FP-lensing signal, as in shear analyses, because they affect the distances inferred from redshifts and make intrinsic correlations between FP residuals more relevant by allowing potential overlap between lens and source samples. These systematics are qualitatively similar to those in traditional shear analyses and can be treated in a similar way \cite{2025MNRAS.537.1924S}. Given the low intrinsic scatter of the FP, it will be exciting to investigate whether photometric FP lensing in Stage-IV surveys can achieve accuracy comparable to, or even better than, shear measurements.
 
Therefore, we conclude that FP lensing provides an accurate and clean probe of cosmic structure, with strong potential for future cosmological galaxy surveys.

\section*{Methods}

\subsection*{Cosmological parameters}
Throughout this paper, we adopt a flat Planck\cite{2020A&A...641A...6P} $\Lambda$CDM cosmology, with $\Omega_{\rm m}=0.3111$ and $H_0=100h\,{\rm km\,s^{-1}\,Mpc^{-1}}$.

\subsection*{DESI galaxy samples}

DESI is a Stage-IV spectroscopic survey designed to map the large-scale structure of the Universe with spectra for roughly 63 million extragalactic sources over an eight-year programme \cite{2013arXiv1308.0847L,2016arXiv161100036D,2016arXiv161100037D,2022AJ....164..207D}. It uses a 5000-fibre multi-object spectrograph on the 4-m Mayall Telescope at Kitt Peak National Observatory, covers more than $17,000~\rm deg^2$, and operates over $3600$--$9800\,\text{\AA}$ \cite{2016arXiv161100037D,2022AJ....164..207D,2023AJ....165....9S,2024AJ....168...95M,2024AJ....168..245P}. DESI target selection is based on DR9 of the DESI Legacy Imaging Surveys \cite{2017PASP..129f4101Z,2019AJ....157..168D}, combining optical $grz$ imaging from DECaLS \cite{2019AJ....157..168D}, BASS \cite{2017PASP..129f4101Z} and MzLS \cite{2018PASP..130h5001Z}. The targeting, spectroscopic reduction and survey-validation pipelines are described in Refs.~\cite{2023AJ....165...50M,2023AJ....165..144G,2023AJ....166..259S}, and DESI has already delivered competitive cosmological constraints, including recent measurements of dark energy \cite{2025JCAP...07..028A,2025PhRvD.112h3515A}.

\begin{figure}[ht]
\centering
\includegraphics[width=0.8\linewidth]{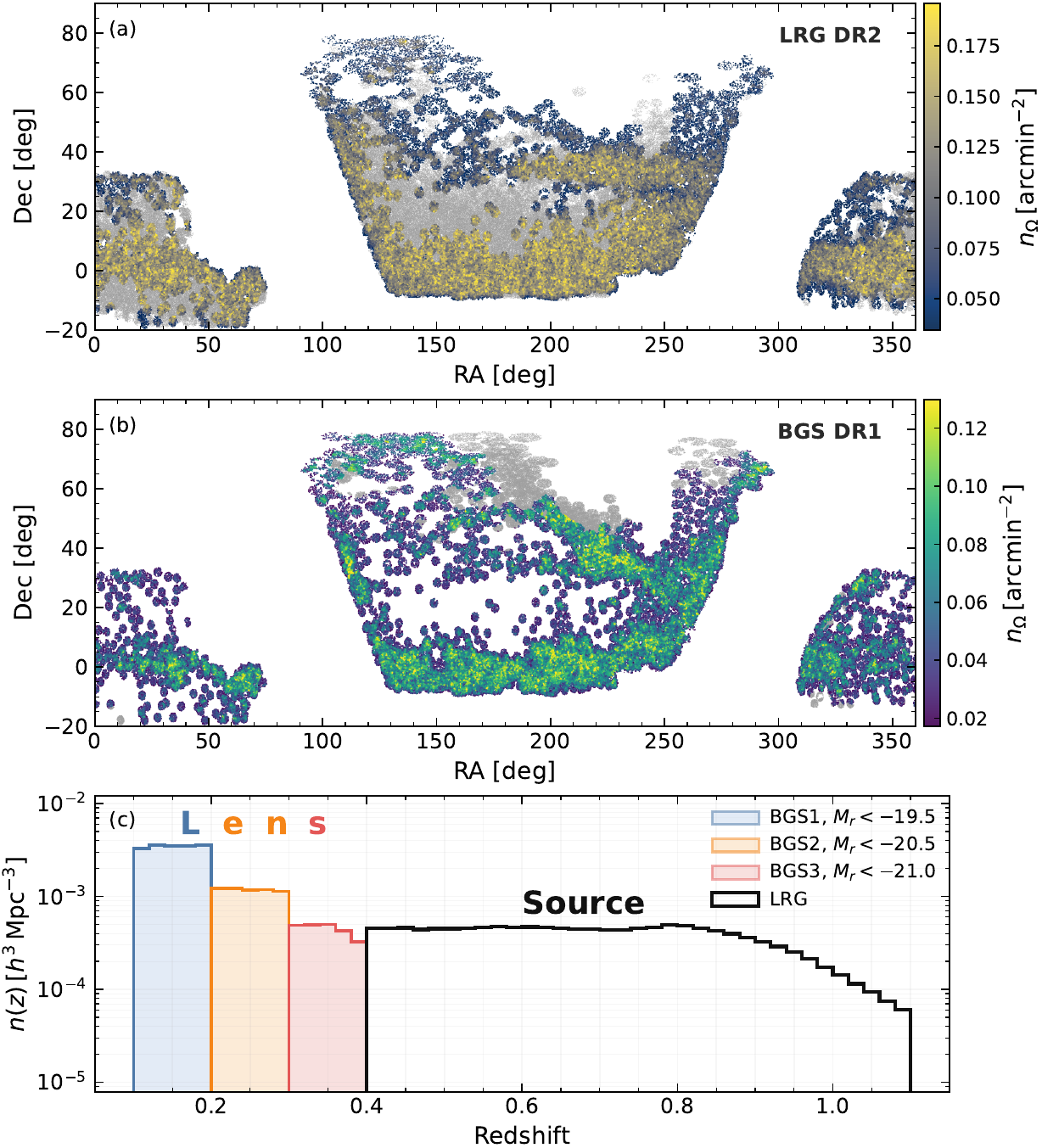}
\caption{
DESI lens and source samples used in this work.
(a) Angular distribution of the DESI DR2 LRG source sample. 
(b) Angular distribution of the DESI DR1 BGS Bright lens sample. 
In panels (a) and (b), the coloured regions show the number density $n_\Omega$ in the overlapping LRG--BGS footprint, while the grey regions indicate the non-overlapping part of the corresponding survey footprint.
(c) Comoving number-density distributions of the three BGS lens samples and the LRG source sample. The lens samples BGS1, BGS2 and BGS3 are defined in the redshift bins $0.1<z<0.2$, $0.2<z<0.3$ and $0.3<z<0.4$, with absolute-magnitude cuts $M_r<-19.5$, $M_r<-20.5$ and $M_r<-21.0$, respectively. The LRG source sample spans $0.4<z<1.1$. For BGS3 lenses, we further require LRG sources to satisfy $z>0.5$.}

\label{fig:dis}
\end{figure}

For the lens sample, we adopt the DESI DR1 BGS Bright lens samples constructed for recent DESI galaxy--galaxy lensing analyses \cite{2024MNRAS.533..589Y,2025arXiv250621677H}. BGS is a core DESI bright-time programme targeting low-redshift galaxies, primarily at $z<0.6$, over a footprint of approximately $17{,}000~\mathrm{deg^2}$ \cite{2023AJ....165..253H}. Its target selection includes a Bright sample with $r<19.5$ and a Faint sample with $19.5<r<20.175$, containing about 10 and 5 million galaxies, respectively; detailed selection criteria and completeness are given in Ref.~\cite{2023AJ....165..253H}. The DR1 BGS Bright sample is based on the first year of DESI main-survey observations \cite{2026AJ....171..285D}. It covers $5300~\mathrm{deg^2}$ in the Northern Galactic Cap and $2173~\mathrm{deg^2}$ in the Southern Galactic Cap, as shown in Figure~\ref{fig:dis}(b), with an average fibre-assignment completeness of 0.656.

The adopted lens catalogue is divided into three redshift bins, $[0.1,0.2]$, $[0.2,0.3]$ and $[0.3,0.4]$, with $r$-band absolute-magnitude cuts of $M_r<-19.5$, $M_r<-20.5$ and $M_r<-21.0$, respectively. These selections were designed in the DESI galaxy--galaxy lensing analyses to maintain an approximately constant number density within each redshift bin. The resulting lens samples are denoted BGS1, BGS2 and BGS3, and are shown in Figure~\ref{fig:dis}(c). After restricting to the $\simeq7264~\mathrm{deg}^2$ overlap with the source sample, the three lens samples contain approximately 739k, 662k and 424k galaxies, respectively.

For the source sample, we use LRGs from DESI DR2. LRGs are a primary DESI dark-time tracer, selected to provide a dense sample of massive galaxies over $0.4<z\lesssim1.1$ for large-scale-structure measurements \cite{2023AJ....165...58Z}. The LRG target selection uses optical $grz$ imaging from the DESI Legacy Imaging Surveys together with WISE\cite{2010AJ....140.1868W} $W1$ photometry, and was designed for high target density, high redshift efficiency and low stellar contamination \cite{2023AJ....165...58Z}. We use the DESI DR2 LRG spectra together with the imaging and spectrophotometric quantities required for the Fundamental Plane analysis. Stellar velocity dispersions, rest-frame photometric quantities and stellar-population measurements are taken from the DESI \textsc{FastSpecFit} catalogue\cite{2023ascl.soft08005M} , which jointly models DESI optical spectra and broadband photometry. Galaxy sizes and shapes are measured by \textsc{Tractor} \cite{2016ascl.soft04008L}. The resulting source sample spans $0.4<z<1.1$; its angular and redshift distributions are shown in Figure~\ref{fig:dis}(b) and Figure~\ref{fig:dis}(c), respectively. The LRG source sample contains around 3.26 million galaxies within the BGS lens footprint.

\subsection*{Fitting the Fundamental-Plane-like relations}

The standard FP defines a tight empirical relation among the effective radius $R_\mathrm{e}$, the central velocity dispersion $\sigma_0$ and the mean surface brightness $I_\mathrm{e}$ within $R_\mathrm{e}$:
\begin{equation}
\log_{10} R_\mathrm{e}
= a \log_{10} \sigma_0 + b \log_{10} I_\mathrm{e} + c\, .
\end{equation}
We convert the angular effective major axis $\theta_\mathrm{e}^a$ measured by \textsc{Tractor} to a physical circularized effective radius in $h^{-1}\mathrm{kpc}$, by setting $R_\mathrm{e} = \theta_\mathrm{e}^a\,D_A\sqrt{b/a}$, where $D_A$ is the angular-diameter distance and $b/a$ is the projected axis ratio. We use this circularized $R_\mathrm{e}$ because shear does not change the half-light area $A_\mathrm{e}=\pi R_\mathrm{e}^2$ at leading order, so $R_\mathrm{e}$ it is invariant to shear at first order. Second-order shear effects on $R_\mathrm{e}$ may be present, but they are expected to be negligible in the weak-lensing regime, although they should be quantified in future work.
We define the mean surface brightness as
\begin{equation}
\log_{10} I_\mathrm{e}
= -0.4 M_r - \log_{10}(2\pi) - 2\log_{10} R_\mathrm{e}\, ,\label{eq:Ie}
\end{equation}
where $M_r$ is the $r$-band absolute magnitude from \textsc{FastSpecFit} with $h=1$. Here we do not further convert $I_\mathrm{e}$ to any specific physical unit, since this choice does not affect our lensing measurement.
The central velocity dispersion $\sigma_0$ is obtained by applying an aperture correction to the velocity dispersion measured by \textsc{FastSpecFit} within the DESI fibre. Following Ref.~\cite{1995MNRAS.276.1341J}, we correct the measured dispersion to an aperture of $R_\mathrm{e}/8$ using
\begin{equation}
\log_{10}\sigma_0
=
\log_{10}\sigma_\mathrm{fib}
+
0.04\log_{10}\left(\frac{\theta_\mathrm{fib}}{\theta_\mathrm{e}/8}\right),
\end{equation}
where $\sigma_\mathrm{fib}$ is the fibre-measured velocity dispersion and $\theta_\mathrm{fib}=0.75''$ is the DESI fibre-aperture radius. 

\begin{figure}[ht]
\centering
\includegraphics[width=0.8\linewidth]{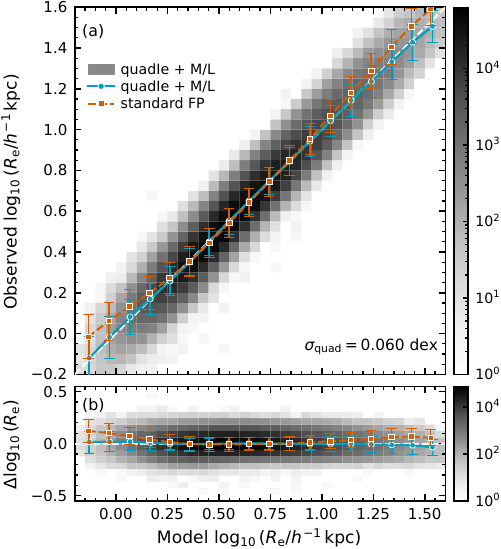}
\caption{
Fundamental-Plane-like size fitting for the DESI LRG source sample.
(a) Observed effective radius versus the radius predicted by the fiducial ``quadIe+ML'' model. The grey-scale histogram shows the two-dimensional number density of galaxies, and the cyan points show the binned mean relation with scatter. The orange squares show the corresponding binned relation for the standard FP fit. The dashed diagonal line indicates the one-to-one relation. The fiducial model gives a residual scatter of $\sigma_{\rm quad}=0.060$ dex.
(b) Residuals in $\log_{10}R_\mathrm{e}$ as a function of the model-predicted size. The ``quadIe+ML'' model removes most of the size-dependent residual trend present in the standard FP and provides a more uniform size prediction across the full range of $R_\mathrm{e}$.
}
\label{fig:FP_fitting}
\end{figure}

The FP can be fitted in several ways, including direct and orthogonal regressions \cite{2012MNRAS.422.1825S}. Here we adopt a direct fit, minimizing the scatter in $R_\mathrm{e}$ at fixed FP observables, because our goal is to predict the intrinsic effective radius of each galaxy from the fitted relation. To account for redshift evolution, we fit the FP in 14 narrow redshift bins with $\Delta z=0.05$ and linearly interpolate the best-fitting parameters across the full redshift range, yielding a model with redshift-dependent coefficients $a(z)$, $b(z)$ and $c(z)$. Figure~\ref{fig:FP_fitting} compares the observed $R_\mathrm{e}$ with the model predictions and shows the mean and scatter of the residuals. Although the standard FP provides an overall good fit with small scatter, it slightly underestimates $R_\mathrm{e}$ for galaxies with both small and large effective radii. 

To obtain a less biased size prediction, we fit an extended FP relation that includes additional lensing-invariant terms: a quadratic dependence on $I_\mathrm{e}$ and a stellar mass-to-light ratio term,
\begin{equation}
\log_{10} R_\mathrm{e}
= a \log_{10} \sigma_0
+ b_1 \log_{10} I_\mathrm{e}
+ b_2 \left(\log_{10} I_\mathrm{e}\right)^2
+ d\log_{10}(M_*/L)
+ c\, ,\label{eq:quadIe}
\end{equation}
where $M_*/L$ (from \textsc{FastSpecFit}) is expressed in units of $M_{\odot}/L_{\odot}$. 
We found this was better than simply working with $I_*\equiv (M_*/L)\,I_e$.
As shown in Figure~\ref{fig:FP_fitting}, this model, labelled ``quadIe+ML'', provides a better fit across the full $R_\mathrm{e}$ range and a slightly smaller scatter, $\sigma_\mathrm{quad}=0.06\,\mathrm{dex}$. The best-fitting parameters of the ``quadIe+ML'' model in each redshift bin are shown in Extended Data Figure~\ref{fig:param}. We find a redshift dependence in the best-fitting parameters of the ``quadIe+ML'' model, mainly driven by the changing LRG selection with redshift. This does not affect our GGL measurement, because any additive bias induced by the fitting procedure is removed by the random-catalogue subtraction described below. We do not show $c$ because we do not use a conventional unit for $I_\mathrm{e}$, which may cause confusion, and the zero point is irrelevant for the lensing measurements.

A further complication is that, by construction, \textsc{FastSpecFit} reports $\sigma_\mathrm{fib}$ only for galaxies for which including the velocity-dispersion parameter significantly improves the spectral fit. In total, 1.37 million LRGs have a reported $\sigma_\mathrm{fib}$ measurement, while 1.89 million do not. The FP-like relations described above are fitted using only galaxies with measured $\sigma_\mathrm{fib}$.

LRGs without $\sigma_\mathrm{fib}$ measurements typically have fainter fibre magnitudes. These galaxies still contain useful information, and removing them would introduce a non-trivial fibre-magnitude selection that would complicate the calculation of the lensing response, as discussed below. To include these objects, we assume in each redshift bin that they follow the same ``quadIe+ML'' relation, but replace the individual velocity dispersion with a fitted mean $\sigma_0$ for the subsample without $\sigma_\mathrm{fib}$. The scatter of the observed $\log_{10}R_\mathrm{e}$ around this relation can then be interpreted as the size-prediction scatter using only $I_\mathrm{e}$ and $M_*/L$. We find that the scatter increases only mildly, to $\sigma_\mathrm{quad}=0.07\,\mathrm{dex}$, because the intrinsic variation of $\sigma_0$ among LRGs is relatively small.

\subsection*{Fundamental-Plane lensing}
Gravitational lensing maps the source-plane position $\boldsymbol{\theta}^{S}$ to the image-plane position $\boldsymbol{\theta}^{I}$. The local lensing distortion is described by the Jacobian matrix
\begin{equation}
    \mathbf{A}
    =
    \frac{\partial \boldsymbol{\theta}^{S}}{\partial \boldsymbol{\theta}^{I}}
    =
    \begin{pmatrix}
        1-\kappa-\gamma_1 & -\gamma_2 \\
        -\gamma_2 & 1-\kappa+\gamma_1
    \end{pmatrix}\,,
\end{equation}
where $\kappa$ is the convergence and $\gamma_1$ and $\gamma_2$ are the two components of the shear, with $\gamma=\gamma_1+i\gamma_2$. The magnification is given by the inverse determinant of this matrix,
\begin{equation}
\mu
=
\left|\frac{\partial \boldsymbol{\theta}^{I}}{\partial \boldsymbol{\theta}^{S}}\right|
=
\frac{1}{(1-\kappa)^2-|\gamma|^2}\,,
\label{eq:mu}
\end{equation}
and describes the amplification of the observed solid angle. Gravitational lensing conserves surface brightness.

FP-like relations provide an opportunity to measure lensing magnification by comparing the observed galaxy size with the intrinsic size predicted from lensing-invariant quantities \cite{2006ApJ...648L..17B,2014ApJ...780L..16H,2015MNRAS.454..478J}. In the ``quadIe+ML'' relation defined in Equation~\ref{eq:quadIe}, $\sigma_0$, $I_\mathrm{e}$ and $M_*/L$ are all invariant under lensing. This relation can therefore be used to infer the intrinsic galaxy size and construct a magnification estimator. For a circularized size, lensing changes the observed effective radius as $R_{\mathrm{e,obs}}=\mu^{1/2}R_{\mathrm{e,int}}$.  The corresponding size residual,
\begin{equation}
    \delta_r(\mu)
    \equiv
    \Delta\log_{10}R_\mathrm{e}(\mu)
    =
    \log_{10}R_{\mathrm{e,obs}}(\mu)-\log_{10}R_{\mathrm{e,int}},
\end{equation}
therefore responds to lensing as
\begin{equation}
    \delta_r(\mu)
    =
    \frac{1}{2}\log_{10}\mu
    \simeq
    \frac{\kappa}{\ln 10}\,,
    \label{eq:response_net}
\end{equation}
where $\mu\simeq1+2\kappa$ in the weak-lensing limit, $\kappa\ll1$ and $|\gamma|\ll1$. In this case, the corresponding convergence-equivalent size noise is $(\ln 10)\,\sigma_\mathrm{quad}\simeq0.138$, substantially smaller than the typical per-component shape noise in shear measurements.

\subsection*{Additional response induced by magnification bias}
However, in contrast to previous work \cite{2006ApJ...648L..17B,2014ApJ...780L..16H,2015MNRAS.454..478J}, we find that the response of $\delta_r$ to lensing is not simply $\kappa/\ln 10$. Instead, an additional response induced by magnification bias is inevitably introduced. 

To see its effect, note that at fixed $I_\mathrm{e}$, Equation~\ref{eq:Ie} gives a purely geometrical relation between size and magnitude,
\begin{equation}
\left.
\frac{\partial \log_{10}R_\mathrm{e}}{\partial M_r}
\right|_{I_\mathrm{e}}
=
-0.2 .
\end{equation}
If the FP observables $\mathbf{x}=(\sigma_0,I_\mathrm{e},M_*/L,\ldots)$ are held fixed, the predicted size is also fixed. Therefore, locally, the variation of the FP residual is set only by the magnitude variation through the size--magnitude relation at fixed surface brightness:
\begin{equation}
\left.
\frac{\partial \delta_r}{\partial M_r}
\right|_{\mathbf{x}}
=
-0.2 .
\end{equation}
In practice, the measured mean residual--magnitude relation is marginalized over the galaxy population at each magnitude,
\begin{equation}
\bar{\delta}_r(M_r)
\equiv
\langle \delta_r\mid M_r\rangle
=
\int d\mathbf{x}\,
p(\mathbf{x}\mid M_r)
\langle \delta_r\mid M_r,\mathbf{x}\rangle .
\end{equation}
The corresponding marginalized slope is
\begin{equation}
\frac{d\bar{\delta}_r}{dM_r}
=
-0.2
+
\int d\mathbf{x}\,
\frac{\partial p(\mathbf{x}\mid M_r)}{\partial M_r}
\langle \delta_r\mid M_r,\mathbf{x}\rangle .\label{eq:slope}
\end{equation}
Thus the marginalized residual--magnitude slope is usually shallower than $-0.2$ because the distribution of position on the plane, $p(\mathbf{x}\mid M_r)$, varies with magnitude, but it generally remains non-zero. If the marginalized residual--magnitude relation is globally linear, $\bar{\delta}_r=\alpha M_r+\beta$, and the least-squares residual satisfies the orthogonality conditions $\mathrm{Cov}(\delta_r,\log_{10}I_\mathrm{e})=0$ and $\mathrm{Cov}(\delta_r,\log_{10}R_{\mathrm{e}}^{\mathrm{fit}})=0$, Equation~\ref{eq:slope} is equivalent to
\begin{equation}
\frac{d\bar{\delta}_r}{dM_r}
=
\frac{\mathrm{Cov}(\delta_r,M_r)}{\mathrm{Var}(M_r)}
=
-5\,\frac{\mathrm{Var}(\delta_r)}{\mathrm{Var}(M_r)}\,.\label{eq:slope_est}
\end{equation}
Figure~\ref{fig:response}(a) shows the dependence of $\bar{\delta}_r$ on $M_r$ for a representative redshift bin. The measured slope is about $-0.095$, close to the value $-0.090$ estimated from Equation~\ref{eq:slope_est}. The remaining difference may arise from departures from a globally linear residual--magnitude relation and from imperfect orthogonality of the fitted residuals.

Galaxy selection often includes a magnitude cut, so magnification can bring intrinsically fainter galaxies into the observed sample. This selection effect is usually referred to as magnification bias. If the newly included fainter galaxies followed the same mean residual distribution as the original sample, the response of $\delta_r$ would remain $\kappa/\ln 10$. However, because $\bar{\delta}_r(M_r)$ has the negative slope shown above, the newly included fainter population has a smaller mean residual than the original sample. Averaging over the selected sample therefore reduces the net response relative to $\kappa/\ln 10$. The effect has the opposite sign in demagnified regions, where galaxies near the selection boundary are removed from the sample. 

Fortunately, this additional response can be estimated directly from the sample. For a magnitude-limited sample, the net response can be written as
\begin{align}
    \langle\delta_r\rangle(\kappa) = \frac{\mathcal{R}\kappa}{\ln 10}\,,\qquad
    \mathcal{R} &= 1 + 5\ln 10\,s(m_\mathrm{lim})
    \left[\bar{\delta}_r(m_\mathrm{lim})-\langle\delta_r\rangle(0)\right]\,,\label{eq:res_mag}
\end{align}
where
\begin{equation}
    s(m_\mathrm{lim})
    \equiv
    \left.
    \frac{d\log_{10}N(<m)}{dm}
    \right|_{m=m_\mathrm{lim}}
\end{equation}
is the logarithmic slope of the cumulative number counts at the limiting magnitude, and $\bar{\delta}_r(m_\mathrm{lim})-\langle\delta_r\rangle(0)$ is the residual offset of galaxies entering or leaving the sample relative to the sample mean. For a perfectly calibrated FP-like model, $\langle\delta_r\rangle(0)=0$.

More generally, the selection depends on multiple total and fibre magnitudes in different bands, which we denote by the vector $\mathbf{m}=(m_1,m_2,\ldots)$. For a selection function $S(\mathbf{m})$, the sample-averaged residual is
\begin{equation}
\langle\delta_r\rangle_S
=
\frac{
\int d\mathbf{m}\, n(\mathbf{m})S(\mathbf{m})\bar{\delta}_r(\mathbf{m})
}{
\int d\mathbf{m}\, n(\mathbf{m})S(\mathbf{m})
}\, ,
\end{equation}
where $n(\mathbf{m})$ is the number density in magnitude space and $\bar{\delta}_r(\mathbf{m})$ is the mean residual at fixed $\mathbf{m}$. Under magnification, a galaxy with unlensed magnitude vector $\mathbf{m}$ is
observed at $\mathbf{m}+\Delta\mathbf{m}$, where
$\Delta\mathbf{m}=-5\kappa\mathbf{q}/\ln 10$ in the weak-lensing limit.
Thus, when integrating over the same intrinsic galaxy population, the
selection function changes from $S(\mathbf{m})$ to
$S(\mathbf{m}+\Delta\mathbf{m})$. Including the direct size response, the
mean residual becomes
\begin{equation}
\left\langle\delta _r\right\rangle_S(\kappa)
=
\frac{\kappa}{\ln 10}
+
\frac{
\int d\mathbf{m}\,n(\mathbf{m})
S(\mathbf{m}+\Delta\mathbf{m})\,
\bar{\delta}_r(\mathbf{m})
}{
\int d\mathbf{m}\,n(\mathbf{m})
S(\mathbf{m}+\Delta\mathbf{m})
}.
\end{equation}
To first order in $\kappa$,
$S(\mathbf{m}+\Delta\mathbf{m})
\simeq S(\mathbf{m})
-(5\kappa/\ln 10)\,
\mathbf{q}\cdot\nabla_{\mathbf{m}}S(\mathbf{m})$.
Expanding the numerator and denominator to first order then gives
\begin{equation}
\mathcal{R}
=
1
-
5
\frac{
\int d\mathbf{m}\, n(\mathbf{m})
\left[\mathbf{q}\cdot\nabla_{\mathbf{m}}S(\mathbf{m})\right]
\left[\bar{\delta}_r(\mathbf{m})-\langle\delta_r\rangle_S\right]
}{
\int d\mathbf{m}\, n(\mathbf{m})S(\mathbf{m})
}\, .
\end{equation}
Here $\mathbf{q}$ allows different magnitude definitions, such as total and fibre magnitudes, to have different responses to lensing.

For DESI LRGs, the selections most relevant for magnification bias are the $z$-band fibre-magnitude cut, $z_\mathrm{fiber}$, and the $r-W1$ colour-dependent $W1$-band magnitude cut \cite{2023AJ....165...58Z}. For $z_\mathrm{fiber}$, we adopt the lensing response factor $q$ from Ref.~\cite{2023JCAP...11..097Z}, which depends on galaxy morphology, $R_\mathrm{e}$ and $b/a$. We then apply a small demagnification, $d\kappa=-0.003$, to the catalogue and measure the corresponding response factor $\mathcal{R}$. Figure~\ref{fig:response}(b) shows $\mathcal{R}$ in different redshift bins, with uncertainties estimated by jackknife resampling. We find $\mathcal{R}\simeq0.3$--$0.5$ for DESI LRGs. We also test different values of $d\kappa$, from $-0.001$ to $-0.005$, in Extended Data Figure~\ref{fig:response_compare_dk}. The choice of $d\kappa$ reflects a balance between Poisson noise, which favours larger $|d\kappa|$, and locality of the response estimate, which favours smaller $|d\kappa|$. We find that the inferred $\mathcal{R}$ values are consistent within the uncertainties for these choices, and adopt $d\kappa=-0.003$ as a good compromise.

\subsection*{A "Eulerian" view of the additional response}
Although the expression above provides a practical way to estimate $\mathcal{R}$ from the sample, a ``Eulerian'' view, in which the response is analysed at fixed magnitude, gives an alternative expression in terms of more elementary quantities: the residual--magnitude relation and the shape of the number counts. This form offers additional insight into the origin of the response. 

For a magnitude-limited sample with observed selection $m<m_{\rm lim}$, let
$n_0(m)={\rm d}N_0/{\rm d}m$ be the unlensed differential number count,
$p_0(m)=n_0(m)/N_0$ the normalized magnitude distribution of the unlensed selected sample, and
$h_0(m)\equiv {\rm d}\ln n_0/{\rm d}m$, where
$N_0=\int^{m_{\rm lim}} n_0(m)\,{\rm d}m$.
Under weak magnification, a galaxy observed at magnitude $m$ is drawn from an intrinsically fainter magnitude,
$m_{\rm int}=m+5\kappa/\ln10$, so the lensed number count is, to first order,
$n_\kappa(m)=n_0(m+5\kappa/\ln10)$.
The normalized lensed distribution is $p_\kappa(m)=n_\kappa(m)/N_\kappa$, with
$N_\kappa=\int^{m_{\rm lim}} n_\kappa(m)\,{\rm d}m$. 

Then, at fixed observed magnitude, the response of $\delta_r$ to magnification can be split into three parts. Consider the case in which the marginalized residual--magnitude relation is globally linear, $\bar{\delta}_r(m)=\alpha m+\beta$. The direct size response contributes $\kappa/\ln 10$. The magnification-induced magnitude shift changes the local mean residual, because galaxies observed at a fixed magnitude are drawn from intrinsically fainter magnitudes, contributing $5\alpha\kappa/\ln 10$. Finally, magnitude shift also changes the number density and therefore the normalized weights $p_\kappa(m)$, producing an additional change in the global mean residual,
\begin{align}
\left.
\frac{\partial \ln p_\kappa(m)}{\partial \kappa}
\right|_{\kappa=0}
&=
\frac{5}{\ln 10}
\left[
\frac{d\ln n_0(m)}{dm}
-
\int dm'\,p_0(m')\frac{d\ln n_0(m')}{dm'}
\right]
\nonumber\\
&=
\frac{5}{\ln 10}
\left[
h_0(m)-\langle h_0\rangle_{p_0}
\right],
\end{align}
where
$\langle h_0\rangle_{p_0}\equiv \int^{m_{\rm lim}} h_0(m)p_0(m)\,{\rm d}m$
denotes the average over the normalized unlensed magnitude distribution. The corresponding change in the global mean residual from this reweighting is
\begin{equation}
\left.
\frac{d\langle\delta_r\rangle}{d\kappa}
\right|_{\rm weight}
=
\int p_0(m)\bar{\delta}_r(m)
\left.
\frac{\partial\ln p_\kappa(m)}{\partial\kappa}
\right|_{\kappa=0}
\,{\rm d}m
=
\frac{5}{\ln10}
{\rm Cov}_{p_0}\left[\bar{\delta}_r,h_0\right],
\end{equation}
where ${\rm Cov}_{p_0}$ denotes the covariance weighted by the normalized unlensed magnitude distribution $p_0(m)$. If $\bar{\delta}_r(m)$ is linear, this becomes
\begin{equation}
\left.
\frac{d\langle\delta_r\rangle}{d\kappa}
\right|_{\rm weight}
=
\frac{5\alpha}{\ln10}
{\rm Cov}_{p_0}(m,h_0).
\end{equation}
 Therefore, defining $\langle\delta_r\rangle(\kappa)=(\mathcal{R}/\ln10)\kappa$, the total response is
\begin{equation}
\mathcal{R}
=
1+5\alpha+5\alpha\,{\rm Cov}_{p_0}(m,h_0).
\end{equation}

This expression is equivalent to Equation~\ref{eq:res_mag}, but recasts the response in an ``Eulerian'' view, where the effect of magnification is evaluated at fixed observed magnitude. This is complementary to the ``Lagrangian'' view above, where the response is obtained by shifting galaxies across the selection boundary. For power-law number counts, $h_0(m)$ is constant and the covariance term vanishes, giving $\mathcal{R}=1+5\alpha$. For the representative redshift bin shown in Figure~\ref{fig:response}(a), where $\alpha\simeq-0.095$, this gives $\mathcal{R}\simeq0.52$. This is slightly higher than the direct estimate shown in Figure~\ref{fig:response}(b), suggesting an additional suppression from the covariance term. In the fixed-$I_\mathrm{e}$ geometric limit, $\alpha=-0.2$,  and $\mathcal{R}=0$ for power-law counts. Therefore, the non-zero response arises from two effects: the magnitude-dependent distribution of galaxies in FP-observable space, $\partial p(\mathbf{x}\mid m)/\partial m\neq0$, which causes $\alpha\neq-0.2$, and the non-zero curvature of the logarithmic number counts.

For a general selection depending on multiple total and fibre magnitudes $\mathbf{m}$, with $\Delta\mathbf{m}=-5\kappa\,\mathbf{q}/\ln 10$, the response can be written as
\begin{equation}
\mathcal{R}
=
1
+
5\left\langle
\mathbf{q}\cdot\nabla_{\mathbf{m}}\bar{\delta}_r(\mathbf{m})
\right\rangle_S
+
5\,{\rm Cov}_S
\left[
\bar{\delta}_r(\mathbf{m}),
\mathbf{q}\cdot\nabla_{\mathbf{m}}\ln n_0(\mathbf{m})
\right],
\end{equation}
where the averages and covariance are evaluated over the selected sample.

\subsection*{Galaxy--galaxy lensing and additive bias}
Compared with detecting the $\kappa\kappa$ auto-correlation, cross-correlating the FP-based convergence field with foreground large-scale structure provides a cleaner measurement and is therefore better suited as a first step. 

If the FP residual contains an additive systematic contribution, we can write
\begin{equation}
\delta_r
=
\frac{\mathcal{R}}{\ln 10}\kappa
+
s
+
\epsilon ,
\label{eq:drFull}
\end{equation}
where $s$ denotes residual additive systematics and $\epsilon$ represents the stochastic FP residual, including both the intrinsic scatter of the FP-like relation and measurement noise. We remove additive contributions associated with the lens survey geometry using a random-lens subtraction,
\begin{equation}
\widehat{\kappa}(R)
=
\left\langle \frac{\ln 10\,\delta_r}{\mathcal{R}}\right\rangle_{\rm L}
-
\left\langle \frac{\ln 10\,\delta_r}{\mathcal{R}}\right\rangle_{\rm R}
=
\kappa(R)
+
\left[
\left\langle \frac{\ln 10\,s}{\mathcal{R}}\right\rangle_{\rm L}
-
\left\langle \frac{\ln 10\,s}{\mathcal{R}}\right\rangle_{\rm R}
\right] .
\end{equation}
Here the first average is taken around the real lenses, while the second is taken around random points with the same angular selection function. The estimator is unbiased if the response-weighted additive systematic has the same mean around real lenses and random points, $\left\langle s/\mathcal{R}\right\rangle_{\rm L}
=
\left\langle s/\mathcal{R}\right\rangle_{\rm R}$.
Thus, provided that the random catalogue traces the lens survey geometry and selection, and that the residual source systematics are not correlated with the foreground lens density, the random subtraction removes additive survey systematics and recovers the mean convergence profile around the lenses. 

In this sense, $\delta_r$ from both the standard FP and the ``quadIe+ML'' model can, in principle, provide unbiased GGL estimates. However, because the survey geometry and selection function are never known perfectly, larger residual systematics in the FP fitting would require a more accurate random catalogue for their subtraction. We therefore adopt the ``quadIe+ML'' model as our fiducial choice. We refer the reader to Ref.~\cite{2025JCAP...01..125R} for details of the random-catalogue construction.

In practice, we implement the estimator as a weighted average over lens--source and random--source pairs,
\begin{equation}
\widehat{\kappa}(R)
=
\frac{\sum_{l,s\in R} w_l w_s\,\ln 10\,\delta_{r,s}/\mathcal{R}_{s}}
     {\sum_{l,s\in R} w_l w_s}
-
\frac{\sum_{r,s\in R} w_r w_s\,\ln 10\,\delta_{r,s}/\mathcal{R}_{s}}
     {\sum_{r,s\in R} w_r w_s}\, .
\end{equation}
Here the quantities $w_l$, $w_r$ and $w_s$ are the lens, random and source weights, respectively.

The convergence profile is related to the projected surface mass density by $\kappa(R)=\Sigma(R)/\Sigma_{\rm crit}(z_l,z_s)$ 
where we use comoving transverse separation \(R\) and comoving surface density. The corresponding comoving critical surface density is
\begin{equation}
\Sigma_{\rm crit}(z_l,z_s)
=
\frac{c^2}{4\pi G}
\frac{D_s}{D_lD_{ls}}
\frac{1}{(1+z_l)^2} ,
\end{equation}
with \(D_l\), \(D_s\) and \(D_{ls}\) denoting angular-diameter distances to the lens, to the source and between the lens and source, respectively. We obtain the projected surface-density profile by applying the same lens--random subtraction with a pair-dependent \(\Sigma_{\rm crit}\),
\begin{equation}
\widehat{\Sigma}(R)
=
\frac{\sum_{l,s\in R} w_l w_s\,
\Sigma_{\rm crit}(z_l,z_s)\,
\ln 10\,\delta_{r,s}/\mathcal{R}_{s}}
     {\sum_{l,s\in R} w_l w_s}
-
\frac{\sum_{r,s\in R} w_r w_s\,
\Sigma_{\rm crit}(z_r,z_s)\,
\ln 10\,\delta_{r,s}/\mathcal{R}_{s}}
     {\sum_{r,s\in R} w_r w_s}\, .
\end{equation}

We measure $\kappa(R)$ and $\Sigma(R)$ for BGS1, BGS2 and BGS3 using the estimators defined above, and estimate the covariance from 200 jackknife regions. For the BGS lens galaxies, we use the total weight
$w_\mathrm{tot}=w_\mathrm{com}w_\mathrm{zfail}w_\mathrm{imsys}$,
which accounts for fibre-assignment incompleteness, redshift failures and imaging systematics, respectively\cite{2025JCAP...01..125R}. We then randomly assign weight--redshift pairs from the lens sample to the random catalogue, ensuring that the randoms reproduce the lens weighted redshift distribution. For BGS3, we further require LRG sources to satisfy $z>0.5$ to ensure sufficient separation between the lens and source redshifts. The resulting profiles are shown in Figure~\ref{fig:GGL}. 

For comparison with shear-based GGL measurements, we convert our surface-density profile to the excess surface density,
\begin{equation}
\Delta\Sigma(R)
=
\overline{\Sigma}(<R)-\Sigma(R),
\qquad
\overline{\Sigma}(<R)
=
\frac{2}{R^2}\int_0^R \Sigma(R')R'\,{\rm d}R' .
\end{equation}
Shear-based measurements directly constrain $\Delta\Sigma(R)$ through $\gamma_t(R)=\Delta\Sigma(R)/\Sigma_{\rm crit}$, whereas our FP-size magnification measurement first estimates $\Sigma(R)$ from convergence and then applies the above integral transformation. We compare the resulting $\Delta\Sigma(R)$ profile with DES, KiDS and HSC shear-source measurements\cite{2025arXiv250621677H} in Figure~\ref{fig:compare_to_shear}. For the contribution below the innermost measured radius, we extrapolate $\Sigma(R)$ using a power law fitted to the first six radial bins.

\subsection*{Multiplicative bias from size measurements}
Although additive systematics in the GGL measurement can be effectively removed using the random catalogue, multiplicative bias remains a concern. This is particularly relevant because the seeing of the DESI imaging surveys is comparable to, and in some cases larger than, the angular effective radii $\theta_\mathrm{e}$ of DESI LRGs.

To illustrate the possible multiplicative bias, consider the standard FP written in logarithmic variables,
\begin{equation}
r = a s + b i + c ,
\qquad
r\equiv \log_{10} R_\mathrm{e},\quad
s\equiv \log_{10}\sigma,\quad
i\equiv \log_{10} I_\mathrm{e}=-0.4M-2r+\mathrm{const}.
\end{equation}
Suppose that the measured quantities are related to the true ones by affine transformations,
$r_{\rm obs}=(1+m_r) r+c_r$, $s_{\rm obs}=(1+m_s) s+c_s$ and $M_{\rm obs}=(1+m_M) M+c_M$. The observed surface-brightness variable is therefore
$
i_{\rm obs}
=
-0.4(1+m_M)M-2(1+m_r)r+\mathrm{const}.
$
After refitting the FP with the observed quantities,
$r_{\rm obs}=a_{\rm obs}s_{\rm obs}+b_{\rm obs}i_{\rm obs}+c_{\rm obs}$, the observed surface-brightness coefficient becomes
\begin{equation}
b_{\rm obs}
=
\frac{(1+m_r)b}{(1+m_M)+2b(m_M-m_r)} .
\end{equation}
We define the observed FP residual as
$
\delta_{r,\rm obs}
=
r_{\rm obs}
-a_{\rm obs}s_{\rm obs}
-b_{\rm obs}i_{\rm obs}
-c_{\rm obs}.
$
Under lensing magnification, $\Delta r=(1/2)\log_{10}\mu$, $\Delta M=-2.5\log_{10}\mu$ and $\Delta s=0$, so that
$
\Delta i_{\rm obs}
=
(m_M-m_r)\log_{10}\mu .
$
The lensing-induced FP residual is therefore
\begin{equation}
\widehat{\delta}_{r,\rm lens}
=
\Delta r_{\rm obs}
-
b_{\rm obs}\Delta i_{\rm obs}
=
\frac{(1+m_r)(1+m_M)}
{2\,[(1+m_M)+2b(m_M-m_r)]}\log_{10}\mu .
\end{equation}
Relative to the ideal response, $\delta_{r,\rm lens}=(1/2)\log_{10}\mu$, the multiplicative bias is
\begin{equation}
\mathcal{M}
=
\frac{(1+m_r)(1+m_M)}
{(1+m_M)+2b(m_M-m_r)}
-1
=
m_r-2b_{\rm obs}(m_M-m_r)\, .
\label{eq:multibias}
\end{equation}
In weak lensing, this gives $\widehat{\kappa}=(1+\mathcal{M})\kappa$.

Equation~\ref{eq:multibias} shows that the multiplicative bias in $\kappa$ depends on the multiplicative biases in magnitude and size, $m_M$ and $m_r$, and on the surface-brightness coefficient, which can be expressed using either the intrinsic slope $b$ or the observed slope $b_\mathrm{obs}$. Since magnitudes are generally well measured, we expect $m_M\simeq0$, so the dominant contribution comes from size bias. For our standard FP fit, $b_\mathrm{obs}\simeq-0.6$, giving $\mathcal{M}\simeq-0.2m_r$. The resulting bias is therefore modest: even for a large size bias of $m_r=-0.5$, the induced multiplicative bias is only $\mathcal{M}\simeq0.1$. Extended Data Figure~\ref{fig:Re_multibias}(a) shows $\mathcal{M}$ for different values of $b$, $b_\mathrm{obs}$ and $m_r$ with $m_M=0$. If $M_*/L$ is included, it behaves similarly to $\sigma_0$ and does not affect the $\kappa$ measurements.

Although Equation~\ref{eq:multibias} is derived for the standard FP, the ``quadIe+ML'' model differs only mildly from the standard FP in the fitted size relation, as shown in Figure~\ref{fig:FP_fitting}. We therefore expect the multiplicative bias of the ``quadIe+ML'' model to be close to the standard-FP prediction. Extended Data Figure~\ref{fig:Re_multibias}(b) provides a direct test of this expectation for the ``quadIe+ML'' case.

\subsection*{Data availability}
All raw data used in this work will be released as part of DESI Data Release~2. The data points shown in each figure are available in machine-readable form at \url{https://doi.org/10.5281/zenodo.22311386}. Additional derived data products are available from the corresponding author upon reasonable request.

\bibliography{sample}

\section*{Acknowledgements}
We thank Pengjie Zhang and Yan Lai for useful comments during the DESI Collaboration Wide Review, and Yipeng Jing and Zhenjie Liu for helpful discussions. K.X. is supported by funding from the Center for Particle Cosmology at U Penn. R.K.S. is grateful to the ICTP and IFPU in Trieste for their hospitality when this work was completed.  This work made use of the Gravity Supercomputer at the Department of Astronomy, Shanghai Jiao Tong University.

This material is based upon work supported by the U.S. Department of Energy (DOE), Office of Science, Office of High-Energy Physics, under Contract No. DE–AC02–05CH11231, and by the National Energy Research Scientific Computing Center, a DOE Office of Science User Facility under the same contract. Additional support for DESI was provided by the U.S. National Science Foundation (NSF), Division of Astronomical Sciences under Contract No. AST-0950945 to the NSF’s National Optical-Infrared Astronomy Research Laboratory; the Science and Technology Facilities Council of the United Kingdom; the Gordon and Betty Moore Foundation; the Heising-Simons Foundation; the French Alternative Energies and Atomic Energy Commission (CEA); the Secretariat of Science, Humanities, Technology and Innovation (SECIHTI) of Mexico; the Ministry of Science, Innovation and Universities of Spain (MICIU/AEI/10.13039/501100011033), and by the DESI Member Institutions: \url{https://www.desi.lbl.gov/collaborating-institutions}. Any opinions, findings, and conclusions or recommendations expressed in this material are those of the author(s) and do not necessarily reflect the views of the U. S. National Science Foundation, the U. S. Department of Energy, or any of the listed funding agencies.

The authors are honored to be permitted to conduct scientific research on I'oligam Du'ag (Kitt Peak), a mountain with particular significance to the Tohono O’odham Nation.

\section*{Author contributions}

K.X. performed the data analysis and theoretical interpretation, and wrote the initial draft. R.K.S. proposed the idea. J.W. contributed to the shear-lensing comparison. The other authors contributed to the DESI survey, including its construction, operations, observations, data reduction, validation and data products. All authors reviewed and commented on the manuscript.

\section*{Competing interests}

The authors declare that they have no competing interests.

\section*{Additional information}

\textbf{Correspondence and requests for materials} should be addressed to Kun Xu (kunxu@sas.upenn.edu)

\clearpage
\setcounter{figure}{0}
\renewcommand{\figurename}{Extended Data Figure}

\begin{figure}[ht]
\centering
\includegraphics[width=0.5\linewidth]{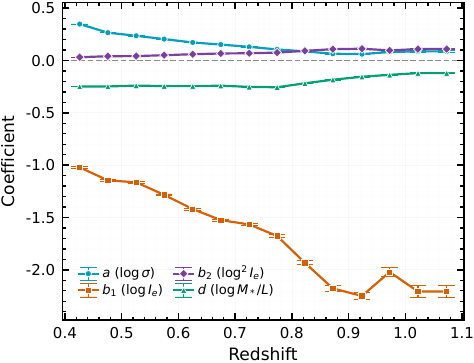}
\caption{
Best-fitting coefficients of the fiducial ``quadIe+ML'' size relation as a function of redshift.
The model includes linear terms in $\log_{10}\sigma$, $\log_{10} I_\mathrm{e}$ and $\log_{10}(M_\ast/L)$, together with a quadratic term in $\log_{10} I_\mathrm{e}$. The points show the coefficients fitted independently in each redshift bin. The redshift evolution of these coefficients is used to predict the intrinsic effective radius of each LRG source.
}
\label{fig:param}
\end{figure}

\begin{figure}[ht]
\centering
\includegraphics[width=0.5\linewidth]{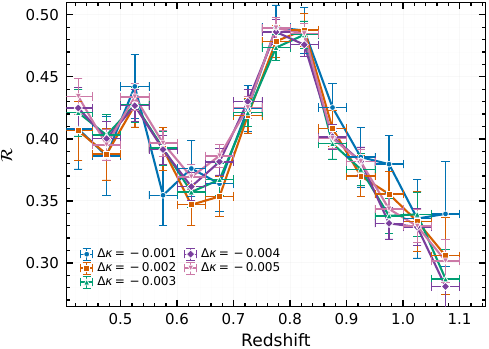}
\caption{
Test of the response calibration using different applied convergence values.
The effective lensing response $\mathcal{R}$ is measured as a function of source redshift after applying small test convergences, $\Delta\kappa=-0.001$ to $-0.005$, to the LRG source sample. The consistency among the different $\Delta\kappa$ values shows that the response estimate is stable in the weak-lensing regime. Error bars are estimated from jackknife resampling.
}
\label{fig:response_compare_dk}
\end{figure}

\begin{figure}[ht]
\centering
\includegraphics[width=0.6\linewidth]{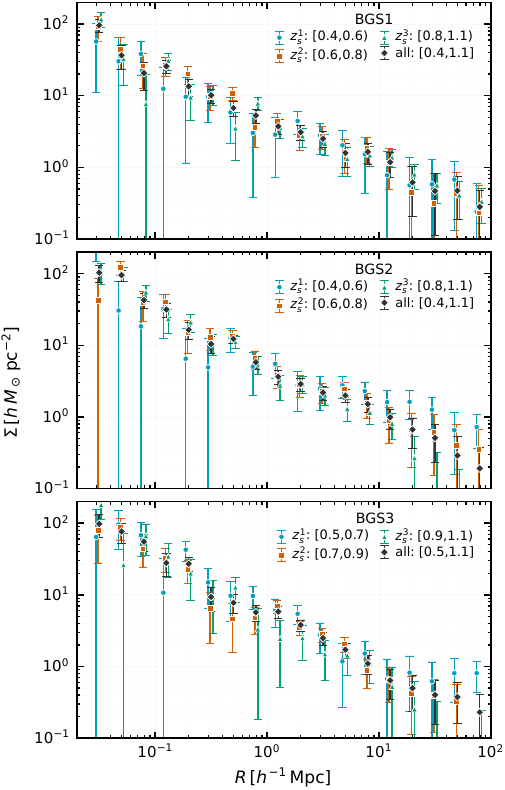}
\caption{
Tomographic consistency of the FP-lensing surface-density measurements.
Panels show the surface-density profiles $\Sigma(R)$ for the BGS1, BGS2 and BGS3 lens samples, measured using different source-redshift bins. The coloured points show measurements from individual LRG source bins, while the black diamonds show the measurement using the full source sample. The consistency among the source bins indicates that the signal is stable against changes in the source-redshift selection. All measurements are in comoving coordinates, and error bars show jackknife uncertainties.
}
\label{fig:GGL_Sigma_sourcebins}
\end{figure}

\begin{figure}[ht]
\centering
\includegraphics[width=0.8\linewidth]{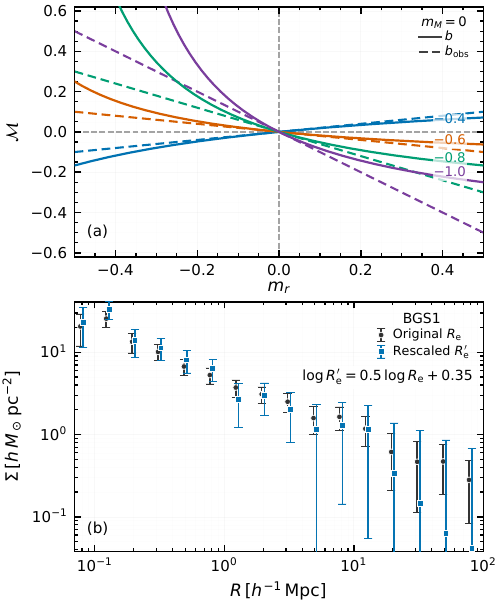}
\caption{
Impact of size multiplicative bias on the FP-lensing measurement.
(a) Multiplicative bias $\mathcal{M}$ in the inferred convergence as a function of size bias $m_r$, assuming $m_M=0$. Solid curves show the prediction using the intrinsic FP surface-brightness slope $b$, while dashed curves show the equivalent expression using the observed slope $b_{\rm obs}$. The labelled curves correspond to different values of the FP slope.
(b) Direct test using the BGS1 lens sample. The black points show the fiducial $\Sigma(R)$ measurement using the original effective radii, while the blue squares show the result after applying a deliberately large size rescaling, $\log_{10} R_\mathrm{e}'=0.5\log_{10} R_\mathrm{e}+0.35$, and refitting the fiducial ``quadIe+ML'' model. The consistency of the two measurements demonstrates the weak sensitivity of the FP-lensing signal to size multiplicative bias.
}
\label{fig:Re_multibias}
\end{figure}

\end{document}